\documentclass[12pt]{article}  
\newcommand{\comma}{\;\; ,}
\newcommand{\period}{\;\; .}

\newcommand{\eq}{\; = \;}
\newcommand{\sep}{\;\; , \;\;}
\newcommand{\be}{\begin{equation}}
\newcommand{\bd}{\begin{displaymath}}
\newcommand{\ee}{\end{equation}}
\newcommand{\ed}{\end{displaymath}}
\newcommand{\ba}{\begin{eqnarray}}
\newcommand{\ea}{\end{eqnarray}}

\newcommand{\taub}{\widehat{\tau}}

\newcommand{\minus}{\! - \!}

\newcommand{\om}{\omega}
\newcommand{\spc}{\; \; \; \;}

\title{The  $\tau_2$ model and parafermions}

\author{ R.J. Baxter\\
{\protect \small Mathematical
Sciences Institute}\\
{\protect \small  The Australian National University,
 Canberra, A.C.T. 0200, Australia  }
}

\date{\small Monday 17 March 2014}

\begin{document}

%\magnification = \magstep1
%\magnification = 1000

\maketitle

\abstract{Paul Fendley has recently found a ``parafermionic'' way to diagonalise a 
simple solvable hamiltonian associated with the chiral Potts model. Here we 
indicate how this method  generalizes to the $\tau_2$ model with open boundaries and make some comments.}

 %%34567890123456789012345678901234567890123456789012345678901234567890

\section{Introduction}
\setcounter{equation}{0}
For a given integer $N$, let
\be \omega =  { \rm e}^{2 \pi {\rm i} /N} \period  \ee
Let $I,  Z,  X$ be   $N$-by-$N$ matrices, $I$ the identity and $Z, X$  having elements
\be Z_{jk} \eq \om^{j-1} \delta_{jk}  \sep X_{jk} \eq \delta_{j,k+1} \comma \ee
and  $Z_m$ the $N^L$-dimensional matrix
\be   Z_m \eq   I \otimes  \cdots  I  \otimes Z \otimes  I \cdots \otimes I  
\ee
where there are $L$ terms in the direct product and $Z$ is in position $m$. Similarly, let
\be   X_m \eq   I \otimes  \cdots  I  \otimes X \otimes  I \cdots \otimes I  \period
\ee

In 1989\cite{Bax1989,Bax1989a} the author posed a puzzle by showing,{\footnote{ I 
have slightly simplified the puzzle by taking $r = L+1$ and $\gamma_1, \gamma_r =0$ 
in \cite{Bax1989}, then re-labelling $\gamma_2, \ldots, \gamma_{r-1}$
as $\gamma_1, \ldots, \gamma_{L-1}$}}
  via a quite  roundabout route 
 that the hamiltonian
\be \label{hamintro}
 {\cal H} \eq  - \sum_{j=1}^L  \alpha_j \,  X_j   - \sum_{j=1}^{L-1}  \gamma_j
 Z_j Z_{j+1}^{-1}  \ee
 has the same eigenvalues as the diagonal matrix
 \be \label{dirsum}
 {\cal H} \eq   - \sum_{j=1}^L  \zeta_j \,  Z_j  \ee
 
\setlength{\jot}{9mm}
 
where $\zeta_1, \ldots, \zeta_L$ are the solutions of the equation
 \be
 \label{fnD1}
\left| \begin{array}{ccccccc} 
-\zeta^{N/2}   & g_1^{N/2} & 0 & 0 & .. & 0  & 0   \\  
g_1^{N/2}  & -\zeta^{N/2}  &  g_2^{N/2}  & 0 &  ..  &   0  & 0 \\
0 & g_2^{N/2}  & -\zeta^{N/2}  &  g_3^{N/2}    & .. &  0 & 0   \\
0 & .. & .. & .. & .. & .. & 0 \\
  0  & 0 & 0 & 0 & g_{2L-2}^{N/2}   & -\zeta^{N/2}   &  g_{2L-1}^{N/2} \\
 0  & 0  & 0 & 0 & 0 & g_{2L-1}^{N/2}   & -\zeta^{N/2}   
\end{array} \right|   \eq 0  \period \ee

Despite appearances, this determinant is a multinomial in integer powers
of  $\zeta^N, g_1^N, \ldots, g_{2L-1}^N$. It is of degree $L$ in $\zeta^N$. I then 
said that this suggested there might be a simpler way of obtaining this result, 
similar to the spinor operator or Clifford algebra method used by 
Kaufman {Kaufman49} for the Ising model. I repeated this suggestion
in 2004.\cite{Bax04}

This puzzle has now been solved by Paul Fendley, using parafermion 
operators.\cite{Fendley13} Here I wish to 
suggest an extension of Fendley's method  to the more general problem of the 
$\tau_2$ model with open boundaries.\cite{Bax04}

The equations of sections 4, 5 and 6 have not been rigorously proved, but are conjectures
based on algebraic  computer calculations for small values (between 2 and 6)
of $N$ and $L$. Since my original presentation of this work,\cite{Baxter2013}  equations (\ref{relngammas}) and (\ref{numu}) have been proved by Helen Au-Yang and Jacques Perk.\cite{AuYangPerk} Their paper follows this.

 %%34567890123456789012345678901234567890123456789012345678901234567890

\section{$\tau_2$ model}
\setcounter{equation}{0}
The $\tau_2$ model, with weights that vary from column to column,  can be defined \cite{BBP,Bax04} as follows. Consider the square lattice  with $M$ rows of  $L+1$ sites  and
periodic boundary conditions. (In section 3  we effectively remove the end column, which 
is the reason for this choice of notation.) At each site $i$ (numbered $i = 0, 1, 2, \ldots, L$) 
there is  a ``spin'' $\sigma_i$, which takes 
the values $0, \ldots , N-1$. Constrain them so that if $j$ is the site immediately above $j$, 
then

\be \sigma_j  \; \eq  \; \; \sigma_i \;\;\;  {\rm or} \; \; \;   \sigma_i  - 1 \period \ee

Let $\omega = \rm{e}^{2 \pi \imath/N}$ be the primitive $N$th root of unity.  Similarly to 
(15) of \cite{Bax04}, define a function $F_{pq}(a,m)$, for $a,m = 0,1$,  by 

\ba  & F_{pq} ( 0,0) =  b_p  \;  , &  F( 0,1) =  -\omega c_p t_q  \; , \nonumber \\
 &   F_{pq} ( 1,0) = d_p \;  , &  F( 1,1) =  -\omega a_p  \; .\ea

If $a, b, c, d$ are the four spins round a face, arranged as in Fig 1, then to that face assign 
a  Boltzmann weight
\bd  W_{pp'q}(a, b, c, d) \eq \sum_{m=0}^1 \omega^{m(d-b) } \, (-\omega t_q )^{a-d-m} 
\, F_{pq}(a-d,m) \, F_{p' q} (b-c,m) \ed
as in (14) of \cite{Bax04}. Hence
\ba \label{wts}
&& \spc W_{pp'q}(a,b,b,a)  \spc  \; \; \; = \spc b_p b_{p'} - \om^{a-b+1} t_q \, c_p c_{p'} 
\nonumber \\
&& \; \; W_{pp'q}(a,b,b,a \minus 1)
 \spc  =  \; - \,  \om t_q \, d_p b_{p'} + \om^{a-b+1} t_q \, a_p c_{p'} \nonumber  \\
&& \; \; W_{pp'q}(a,b,b \minus 1,a)
\spc =  \spc  b_p d_{p'} - \om^{a-b+1} c_p a_{p'}  \\
&& W_{pp'q}(a,b,b \minus 1,a \minus  1)
\;   =   - \, \om t_q \, d_p d_{p'} + \om^{a-b+1} a_p a_{p'}  \period \nonumber \ea

 %%34567890123456789012345678901234567890123456789012345678901234567890

Take the  parameters $p, p'$ of the face between columns $j$ and $j+1$ to be 
$p = p_{2j-1}$, $p' = p_{2j}$. If $p = p_m$, write
\be  a_p = a_m \sep  b_p = b_m \sep c_p= c_m \sep d_p = d_m  \ee
and similarly for $p'$. Then the weight of the face between columns $j$ and $j+1$ is
\be W_j (\sigma_j,\sigma_{j+1}, \sigma_{j+1}', \sigma_j'  | t_q) \eq 
W_{p_{2j-1}, p_{2j},q} (\sigma_j,\sigma_{j+1}, \sigma_{j+1}', \sigma_j' ) \period \ee

Thus there are $2 L+2$ sets of parameters $ (a_p,b_p,c_p,d_p)$, for  $p = p_{-1},p_0,  
\ldots, p_{2L}$. We do {\em not} impose any constraints (such as (2.4) of \cite{BBP}) on $a_p,b_p,c_p,d_p$. As in \cite{Bax04}, all these $2L+2$ parameters can be chosen 
arbitrarily.

With these definitions,  the  transfer matrix of the $\tau_2$ model is the $N^{L+1}$ by 
$N^{L+1}$ matrix $\tau_2(q)$ with elements
\be  \label{deftau}
[\tau_2(t_q)]_{\sigma, \sigma'}  \eq \prod_{j=0}^{L}  
W_j (\sigma_j,\sigma_{j+1}, \sigma_{j+1}', \sigma_j'  | t_q) \comma \ee
using the cyclic boundary conditions $\sigma_{L+1} =\sigma_0$, 
$\sigma_{L+1}' = \sigma_0'$. 
The RHS is just the product of the face weights of a typical row of the lattice, 
as shown in Fig. 1.

%% The following picture is allowed a space 14cm by 6cm
\begin{figure}[hbt]
\begin{picture}(420,240) (-40,-130)
\multiput(0,0)(80,0){5}{\circle*{5}}
\multiput(0,80)(80,0){5}{\circle*{5}}
\multiput(22,-15)(0,5){23}{.}
\multiput(58,-15)(0,5){23}{.}
\multiput(102,-15)(0,5){23}{.}
\multiput(138,-15)(0,5){23}{.}
\multiput(182,-15)(0,5){23}{.}
\multiput(218,-15)(0,5){23}{.}
\multiput(262,-15)(0,5){23}{.}
\multiput(298,-15)(0,5){23}{.}

\put(0,0) {\line (0,1) {80}}
\put(80,0) {\line (0,1) {80}}
\put(160,0) {\line (0,1) {80}}
\put(240,0) {\line (0,1) {80}}
\put(320,0) {\line (0,1) {80}}

\put(0,0) {\line (1,0) {320}}
\put(0,80) {\line (1,0) {320}}

\put(18,-30) {$p_{-1}$}
\put(58,-30) {$p_0$}
\put(98,-30) {$p_1$}

\put(258,-30) {$p_{2L-1}$}
\put(298,-30) {$p_{2L}$}

\put(-10,32) {$0$}
\put(70,32) {$1$}
\put(229,32) {$L$}
\put(310,32) {$0$}

\multiput(37,-18)(0,6){18}{$<$}
\multiput(277,-18)(0,6){18}{$<$}

\thinlines
\put(3,66) {$\sigma'_0$}
\put(83,66) {$\sigma'_1$} 
\put(-1,-14) {$\sigma_0$}
\put(79,-14) {$\sigma_1$} 
 \put(239,-14){$\sigma_L$}
 \put(243,66){$\sigma'_L$}
\put(314,-14) {$\sigma_{0}$}
\put(323,66) {$\sigma'_{0}$}
\put(340,22) {$\tau_2(t_q)$} 
%\put(-30,12) {1} 
%\put(-30,52) {2}
%\put(-33,132) {$2 M$} 
\put(0,-60) {\small{\bf Figure 1: } \footnotesize{A row of the square lattice 
of $L$ columns, showing }}
\put(0,-76) {  \footnotesize{the spins in the lower and upper rows and the vertical  lines }}
\put(0,-92) { \footnotesize{associated with the 
parameters $p_{-1}, p_0, p_1, \ldots , p_{2L}$.}}
\end{picture}
\end{figure}

 %%34567890123456789012345678901234567890123456789012345678901234567890

We shall also use the $N^{L+1}$-dimensional matrices  $X_j, Z_j$, with entries
\be \label{defZX}
 [Z_j]_{\sigma, \sigma'}  \eq  \om^{\sigma_j} \, 
 \prod_{k=0}^L\delta (\sigma_k,\sigma_k') \sep
 [X_j]_{\sigma, \sigma'}  \eq  \delta(\sigma_j, \sigma_j'+1) \prod_{k \neq j} ^L  
 \delta (\sigma_k,\sigma_k') \ee
 the second product being over all values of $k$ from $0$ to $L$, except $j$. We also 
 take $I$ to be the identity matrix, and set $X = X_0 X_1  X_2 \cdots X_L$.

\subsection*{Functional relations for $\tau_2(t_q)$}
Write $t_q$ simply as  $t$:
\be t_q  \eq t  \period \ee
From the relations of \cite{BBP}, two matrices  $\tau_2(t), \tau_2(t')$, with the same 
parameters $p_{-1}, p_0, \ldots , p_{2L}$ but different $t$, commute:
\be \tau_2(t)  \, \tau_2(t') \eq  \tau_2(t')  \, \tau_2(t)  \period \ee

Define two-by-two matrices  $A_0, A_4, \ldots , A_{2L}$ and $B_{-1}, B_1, $ $ \ldots , 
B_{2L-1}$ by
\be  \label{defAB} A_J  \eq 
\left( \begin{array}{cc} c_J^N \, t^N   & a_J^N \\  b_J^N  & d_J^N \end{array} \right)  
\sep    B_J  \eq 
\left( \begin{array}{cc} - c_J^N   & b_J^N \\ 
 a_J^N  & -d_J^N \, t^N\end{array} \right)  \ee
and set
\be \label{defU}
U \eq B_{-1} A_0 B_1  A_2 \cdots   A_{2L}  \period \ee

 %%34567890123456789012345678901234567890123456789012345678901234567890

Also, from (47) of \cite{Bax04}, one can define related matrices 
$\tau_3(t_q), \ldots , \tau_{N+1}(t_q)$ so that, for   $j = 2, \ldots, N$,
\ba \label{taurelns}
\tau_2(\omega^{j-1} t ) \, \tau_j(t)  & = &  z(\om^{j-1} t) X \tau_{j-1}(t) + \tau_{j+1}(t)  \comma \nonumber \\
\tau_j(\om t ) \, \tau_2(t)  & = &  z(\om t) X \tau_{j-1}(\om^2 t) + \tau_{j+1}(t)  \comma  \\
 \tau_{N+1}(t)  & = &  z( t) X \tau_{N-1}(\om  t) +(\alpha + \overline{\alpha})I   \comma
  \nonumber \ea
 where 
 \be \label{defz}
 \tau_1 (t) = I \sep z(t) \eq \om^L \prod_{J=-1}^{2L} (c_J d_J  - a_J b_J \, t ) \comma \ee
 and $\alpha, \overline{\alpha}$ are the two eigenvalues of $U$, so
 \be \label{sumaa}
 \alpha + \overline{\alpha}   = {\rm Trace} \; U \period \ee
 
Clearly $z(t)$ is a polynomial of degree $2 L+2$.  Each element of the matrix $\tau_2(t)$ 
is a polynomial in $t = t_q$ of degree at most $L+1$.
 It follows that  $\tau_j(t)$ is at most of degree $(j-1)(L+1)$, and this is consistent with the 
 fact that $U,  \alpha + \overline{\alpha}$ are of degree $N (L+1)$.
 
 Since $\tau_2(t)$ commutes with $X$, one can choose a representation where the 
 $\tau_j(t)$ (for all $j$ ) are diagonal, with elements (eigenvalues) that are 
 polynomials in $t$ of degree  $(j-1)(L+1)$. Then (\ref{taurelns}) defines all the 
 eigenvalues of $\tau_2(t), \ldots, \tau_{N+1}(t)$. However, the situation is similar 
 to Bethe ansatz calculations: in general the best one can do is a brute force numerical 
 calculation for each of the $N^(L+1)$ eigenvalues.

 %%   TAU 2 MODEL WITH OPEN BOUNDARIES %%
 
 \section{$\tau_2$ model  with open boundaries}
\setcounter{equation}{0}
 The problem dramatically simplifies if one imposed fixed boundary conditions, instead 
 of cyclic ones. As is remarked in \cite{Bax04}, we can do this easily be taking
 \be \label{setad}
 a_{-1} = d_{-1}= 0 \period \ee
Then for the left-hand weight function $W_0$, the $a_p, d_p$ in (\ref{wts}) are zero. 
Hence $W_0(a,b,c,d) $ vanishes unless $d= a$, i.e. 
\be  \sigma'_0 =\sigma_0 \period \ee

Also,  from (\ref{setad}) and (\ref{defz}),
\be  z(t) = 0 \ee
so the terms in  (\ref{taurelns})  involving $X$ do not occur and all the $\tau_j(t)$ matrices 
are block-diagonal, having non-zero entries only when $  \sigma'_0 =\sigma_0 $. If we also 
choose
\be \label{setcc}
c_{-1}= c_{2L}= 0 \comma \ee
then $  \sigma_0, \sigma'_0 $ no longer enter the RHS of (\ref{deftau}), so all the diagonal 
blocks  are the same and without further loss of generality we can focus on the case
\be  \sigma'_0 =\sigma_0 = 0 \period \ee

 %%34567890123456789012345678901234567890123456789012345678901234567890

This reduces the dimensionality of the transfer matrices:  the $\tau_j(t)$ are  now of 
dimension $N^L$. Hereinafter we take $I$ to be the $N^L$-dimensional identity matrix, 
and re-define  $Z_j, X_j$ to be $N^L$-dimensional matrices given by (\ref{defZX}), the 
products being from $k = 1$ to $k = L$.

We choose the normalization of the weights so that 
\be  \label{beq1}
b_{-1} \; = \; b_0 \; = \; b_1 \; = \; \cdots \; = \; b_{2L} = 1  \ee
which ensures that the allowed values of the end weights $W_0, W_{L} $ are
\ba W_0 (a,b,b,a |t )  =   1 & , &  W_0 (a,b,b-1,a | t)  =   d_0  \nonumber \\
 W_{L} (a,b,b,a | t)  = 1 & , &  W_{L} (a,b,b,a-1 | t)  = -\om \,  t \, d_{2L-1} \ea
 Effectively  each row of the lattice loses the first column of spins and the bordering 
 half-faces,
 leaving the section of Fig.  1 between the zig-zag vertical lines
 
 The relations   (\ref{taurelns})  now greatly simplify. so (\ref{taurelns})   reduces to
\bd \tau_j(t)  \eq  \tau_2(t) \,  \tau_2(\om t)  \cdots  \tau_2(\om^{j-2} t) \ed
\be  \label{taupr}
 \tau_2(t) \,  \tau_2(\om t)   \cdots   \tau_2(\om^{N-1} t) \eq (\alpha +\overline{\alpha} ) I 
 \period \ee
{From} (\ref{defAB}),   
\be  B_{-1}  \eq 
\left( \begin{array}{cc}  0 & 1  \\  0 & 0  \end{array} \right)  \sep A_{2L} \eq 
\left( \begin{array}{cc}  0 & a_{2L}^N   \\  1  &  d_{2L}^N   \end{array} \right)  \comma \ee
so if \be \label{defV}
V  =  A_0 B_1 \cdots  B_{2L-1} \comma \ee
then from  (\ref{defU}),  $U$ is the two-by-two matrix
\be U   \eq 
\left( \begin{array}{cc}  V_{22} & \cdots  \\  0 & 0  \end{array} \right)  \ee
so its two eigenvalues $\alpha, \overline{\alpha}$ can be taken to be
\be \alpha=  V_{22} \sep \overline{\alpha} = 0  \ee
 and (\ref{taupr})  becomes
 \be  \label{tauprd2}
 \tau_2(t) \,  \tau_2(\om t)   \cdots   \tau_2(\om^{N-1} t) \eq   V_{22}  \,  I \period \ee

 %%%   DEFINE  f(t^N)  %%%

The element  $V_{22}$ is a polynomial  in $t^N$ of degree $L$. Write it  as
\be \label{deff}
V_{22} \eq f(t^N) \eq s_0 + s_1 t^N +s_2 t^{2N} + \cdots + s_{L} t^{NL}
 \period \ee
{From}  (\ref{defAB})  and (\ref{beq1}), it readily follows that
\be \label{f0}
f(0) = \eq   s_0  \eq  1  \period \ee
Let $1/r_1, \ldots , 1/r_L$ be the zeros of $f(t^N)$, so
\be \label{eqr}
s_0 \,  r_j^{NL}  + s_1\,   r_j^{N(L-1)}  +s_2 \, r_j^{N(L-2)} + \cdots + s_{L} 
 \eq 0 \comma \ee
then
\be f(t^N) \eq \prod_{j=1}^{L} (1- r_j^N   t^N) \period \ee
 
The matrix  $\tau_2(t)$ is a polynomial of degree $L$, equal to the identity 
matrix when $t=0$:
 \be  \label{tz}
 \tau_2(0)  \eq	  I  \period \ee
Going to the diagonal  representation, it follows from  (\ref{tauprd2}) that 
each eigenvalue
of $\tau_2(t)$ must be of the form
\be \label{res}
[\tau_2(t)]_p \eq \prod_{k=1}^{L} (1 - \om^{p_k+1} r_k t) \comma \ee
where $p_1, \ldots , p_{L}$ are integers with values 
$0, \ldots, N-1$ and $r_1, \ldots , r_L$ are independent of $t$. .
Thus there are  $N^{L}$ distinct possible forms for the eigenvalues. Numerical evidence 
suggests that  there is a one-to-one correspondence between the $N^{L}$ eigenvalues 
of $\tau_2(t)$ and the expressions  (\ref{res}).

Making this assumption, it follows that there is a similarity transformation that takes 
$\tau_2(t)$ to
\be \label{diag}
{\cal{P}} ^{-1} \, \tau_2(t) \,  {\cal{P}  } \eq   \prod_{k=1}^{L} (I - \om r_k  t \,   Z_j )   \comma  \ee
the eigenvector matrix $\cal{P}$ being independent of $t$.

 %%34567890123456789012345678901234567890123456789012345678901234567890

%%   ASSOCIATED HAMILTONIAN H  %%

\subsection{The associated hamiltonian $\cal H$}
We define the associated hamiltonian $\cal H$ to be the coefficient of $ \om t$ in the 
Taylor expansion of $\tau_2(t)$, so
\be  \label{defHH}
\tau_2 (t)  \eq   I   +  \om t {\cal H} + {\rm O} (t^2) \period  \ee
then from  (\ref{wts}) and (\ref{deftau}),
\ba \label{hamH}
 {\cal H}  & = &  -\sum_{j=1}^{L} \, \sum_{k=j}^{L} \, \om^{k-j} d_{2j-2} a_{2j-1}
  \cdots a_{2k-2} d_{2k-1} Z_j\, Z_k^{-1} X_j   \cdots X_k \nonumber  \\
  +  \! \! \! \! & \! \! \! \! & \! \! \! \!    \! \! \! \!  \sum_{j=1}^{L-1} \, \sum_{k=j+1}^{L} \, 
  \om^{k-j} c_{2j-1} a_{2j} \cdots a_{2k-2} d_{2k-1} Z_{j}\, Z_{k}^{-1}  X_{j+1}  
  \cdots X_k \nonumber  \\
 -  \! \! \! \! & \! \! \! \! & \! \! \! \!   \! \! \! \!  \sum_{j=1}^{L-1} \, \sum_{k=j}^{L-1} \,
  \om^{k-j} c_{2j-1} a_{2j} \cdots a_{2k-1} c_{2k} Z_{j}\, Z_{k+1}^{-1}  X_{j+1}  
  \cdots X_{k} \\
 +   \! \! \! \!  & \! \! \! \! & \! \! \! \!   \! \! \! \!   \sum_{j=1}^{L-1} \, \sum_{k=j}^{L-1} \, 
 \om^{k-j} d_{2j-2} a_{2j-1} \cdots a_{2k-1} c_{2k} Z_{j}\, Z_{k+1}^{-1}  X_{j}  
 \cdots X_{k}  \nonumber \ea
 
 {From}  (\ref{diag}) its $N^L$ eigenvalues must be
\be \label{eigH}
-  \sum_{j=1}^L r_j  \om^{p_j} \comma \ee
so  we can write the diagonal form of $\cal H$ as
\be \label{defH}
{\cal H}_d \eq   -  \sum_{j=1}^L r_j \, Z_j \period \ee

%%   A particularly simple case  %%

\subsubsection*{A particularly simple case }
This hamiltonian $\cal H$ simplifies if we further  specialize to the 
case when
\be \label{setaa}
a_1 \; = \; a_2 \; = \; \cdots \; = \; a_{2L-2} = 0  \period \ee 
Define $\alpha_1, \ldots, \alpha_{L},\gamma_1,\ldots, \gamma_{L-1}$ by
\be  \alpha_i = d_{2i-2} d_{2i-1}  \sep  \gamma_i= c_{2i-1} c_{2i} \comma  \ee
then (\ref{hamH}) reduces to 
\be \label{hamH2}
 {\cal H} \eq   - \sum_{j=1}^L  \alpha_j \,  X_j - \sum_{j=1}^{L-1}  \gamma_j
 Z_j Z_{j+1}^{-1}  \comma  \ee
 which is (\ref{hamintro}).

Set \be g_{2j-1}= \alpha_j \sep g_{2j} = \gamma_j \comma \ee
and take $D(t^N)$ to be the   $2L$ by $2L$ tridiagonal matrix:
\be \label{fnD}
D(t^N)  \eq \left( \begin{array}{ccccccc} 1  & 1 & 0 & 0 & .. & 0  & 0   \\  
g_1^N\,  t^N & 1 & 1 & 0 &  ..  &   0  & 0 \\
0 & g_2^N  t^N & 1 & 1  & .. &  0 & 0   \\
0 & .. & .. & .. & .. & .. & 0 \\
  0  & 0 & 0 & 0 & g_{2L-2}^N \,   t^N & 1  & 1 \\
 0  & 0  & 0 & 0 & 0 & g_{2L-1}^N  \,  t^N & 1  
\end{array} \right)   \period \ee
Then  we can write $V_{22}$ as
\be V_{22} = f(t^N) = {\rm det} \, D(t^N)  \ee
and $1/r^N_1, \ldots, 1/r^N_L$ are  the zeros of  $D(t^N)$.
This is the result  mentioned in the Introduction and
obtained in 1989 by a less direct method.\cite[Sec 8]{Bax1989}, \cite{Bax1989a} 

%%34567890123456789012345678901234567890123456789012345678901234567890

 \section{Method using parafermions}
 \setcounter{equation}{0}

 For 
$N=2$ the model is equivalent to the Ising model, whose associated hamiltonian 
also has the sum structure (\ref{eigH}) for its eigenvalues. I 
remarked that it would be interesting to  find a method to obtain  this structure  that 
somehow 
parallels the simple  free-fermion or Clifford algebra method used when $N=2$ by 
Kaufman \cite{Kaufman49}. This has recently been achieved for general $N$ by Paul 
Fendley\cite{Fendley13} for the case when  (\ref{setaa}) is satisfied. Here we 
no longer use this restriction: we allow $a_1, a_2, \ldots, a_{2L-2}$ to be arbitrary.

In Kaufman's method, as  applied to the hamiltonian $\cal H$ of (\ref{hamH}) one 
constructs a set of $2L$ operators $\Gamma_0, \ldots, \Gamma_{2L-1}$ such that the 
commutator of $\cal H$ with each  $\Gamma_j$ is a linear combination of   
$\Gamma_0, \ldots, \Gamma_{2L-1}$.  We
 can take 
\be \Gamma_0 \eq Z_1^{-1} \period \ee
Fendley  found, for  general $N$, that if we generate a set of  matrices $\Gamma_j$
 by successive commutation with $\cal H$,
 then this set closes after $N L$ members. He then went on to obtain relations that give the eigenvalues (\ref{eigH}).

Fendley considered the hamiltonian (\ref{hamH}). Here I shall indicate how 
his method 
can be generalised to the more general  $\tau_2$ model with open 
boundaries discussed 
above, and is associated hamiltonian  $\cal H$ as defined by  (\ref{defHH}). 
I  continue to 
impose  the restrictions (\ref{setad}), (\ref{setcc}), \ref{beq1}), but {\em not} 
(\ref{setaa}). The 
results are based on numerical calculations for small $N$ and $L$ 
(no bigger than 5), so 
are just plausible conjectures.

%%34567890123456789012345678901234567890123456789012345678901234567890

Both $\tau_2(t)$ and $\cal H$ depend on the $6L-4$ independent parameters 
$d_0,  a_1,  c_1,  d_1, $ $ \ldots ,$ $a_{2L-2},$ $c_{2L-2},$ $d_{2L-2}, d_{2L-1}$.
Remarkably, all the following equations will involve these parameters  only 
via the coefficients $s_0, s_1, \ldots, s_L$ in (\ref{deff}). 

\subsection{Relations involving $\cal H$}
 
 Defining $\Gamma_1, \ldots \Gamma_{NL}$ by
\be  \label{defG}
\Gamma_{j+1} \eq   \frac{\om }{1-\om} \,  ( { \cal H } \Gamma_{j} -  \Gamma_{j} 
{ \cal H } ) \; \; \; , \; \; \;  
 j = 0, 1, \ldots  \comma \ee
 we find that
 \be  \label{relngammas}
 s_0  \, \Gamma_{NL} + s_1 \, \Gamma_{N(L-1)}+ s_2 \,   \Gamma_{N(L-2)} + 
 \, \cdots \, + s_L \, \Gamma_0 \eq 0 \period \ee

 \subsection{Relations involving $\tau_2(t)$}

Set
\be   q = NL-1 \period \ee
We find for  all $j$  that  $\tau_2(t) \Gamma_j \tau_2(t)^{-1} $ is a 
 linear combination of $\Gamma_0, \ldots , \Gamma_q$. Set
 \be X \eq X_1 X_2 \cdots X_L \period \ee
 From
  (\ref{wts}) and (\ref{deftau}),  the matrix $\tau_2(t)$ is a polynomial in $t$ of degree 
  $L$,  its coefficient of $t^0$ being $I$ and the coefficient of $t^L$ being 
  $\om  d_0 \cdots d_{2L-1} X^{-1}$. The matrix $\tau_2(t)$ commutes with $X$. 
  Each $\Gamma_j$  is 
  independent of $t$ and satisfies the $\om$-commutation relation
  \be X  \, \Gamma_j \eq \om \Gamma_j  \,  X \ee
  so, for $0 \leq  j   \leq  NL$, 
  \be \label{defmu}  \mu_j \eq \Gamma_{j} \tau_2(t) - \tau_2(t) \Gamma_{j} \ee 
  is a polynomial in $t$ whose coefficient of $t^0$ is zero and 
  \bd  \nu_j \eq \om \Gamma_j \tau_2(t) - \tau_2(t) \Gamma_j  \ed 
    is a polynomial in $t$ whose coefficient of $t^L$ is zero. We observe numerically that
    \be  \label{numu}
     t \,  \nu_j  \eq  \mu_{j-1}  \ee
 for  $0 < j  \leq NL $, and find an equation very similar to (\ref{relngammas}), namely
 \be  \label{relnnus}
 s_0  \, \nu_{NL} + s_1 \, \nu_{N(L-1)}+ s_2 \,   \nu_{N(L-2)} + 
 \, \cdots \, + s_L \, \nu_0 \eq 0 \period  \ee

Let $H$ be the $NL$ by $NL$ matrix, with elements $h_{jk}$ and  rows and columns 
labelled $0, \ldots, q$:
 \be  \label{matH}
 H  \eq  \left( \begin{array}{cccccccc} 0 & 1 & 0 & 0 & .. & 0 & 0  & 0   \\  
0  & 0 & 1 & 0 &  ..  &   0 & 0  & 0 \\
0 & 0 & 0 & 1  & .. &  0 & 0 & 0   \\
0 & .. & .. & .. & .. & .. & .. & 0 \\
  0  & 0 & 0 & 0 & ..  & 0 &  0  & 1 \\
  - s_L  & 0  & 0 & -s_{L-1}  & ..  & -s_1 &   0 & 0  
\end{array} \right)    \period \ee
(The last row is shown for the case $N = 3$: in general there are $N-1$ zeros 
between  $-s_j$ and $-s_{j-1}$, and after $-s_1$.) 

Then (\ref{defG}),
 (\ref{relngammas})  are equivalent to:
\be \label{Heqn}
\Gamma_j {\cal H}  - {\cal H} \Gamma_j  \eq (1 - \om^{-1}) 
\sum_{k=0}^{NL-1} h_{jk} \, \Gamma_k  \ee
and (\ref{defmu}), (\ref{relnnus}) to
\be \mu_j  \eq  t
\sum_{k=0}^qh_{jk} \, \nu_k  \comma \ee
i.e.
\be \label{Heqn2}
\Gamma_{j} \tau_2(t) - \tau_2(t) \Gamma_{j}  \eq  t
\sum_{k=0}^q  h_{jk} \, [  \om \Gamma_k \tau_2(t) - \tau_2(t) \Gamma_k  ]   \ee
for $ 0 \leq j  < NL$.
If we define the $NL$ by $NL$ matrix 
\be M = (I-tH)^{-1} (I-\om t H)  \comma \ee
then (\ref{Heqn2}) can be written
\be  \label{treps}
\tau_2(t) \, \Gamma_j  \,  \tau_2(t)^{-1} \eq \sum_{k=0}^q   M_{jk} \, \Gamma_k 
\period \ee
Thus $M$ is the representative of $\tau_2(t)$, in the sense of the 
Kaufman-Onsager representative matrices for the Ising model.\cite{Bax77}

\subsection{The eigenvalues of $\cal H$, $\tau_2(t)$}

Let $P$ be the $N L$ by $N L$ matrix that 
diagonalizes $H$:
\be H  P \eq P  H_d \comma \ee
where $H_d$ is the diagonal matrix with elements 
$\lambda_1, \ldots , \lambda_{NL}$.
Set
\be \label{lintr}
\widehat {\Gamma}_i  \eq \sum_{j=0}^q (P^{-1})_{ij} \, \Gamma_j \period \ee
Multiplying (\ref{Heqn}) by $(P^{-1})_{ij}$ and summing over $j$, it becomes
\be \label{eqsh}
\widehat{ \Gamma}_i {\cal H}  - {\cal H} \widehat {\Gamma}_i  \eq (1 - \om^{-1}) 
\lambda_i \,  \widehat {\Gamma}_i  \period \ee
Similarly, (\ref{Heqn2})  becomes
\be   \label{eqst} \widehat {\Gamma}_{i} \tau_2(t) - \tau_2(t) \widehat {\Gamma}_{i}  \eq  t  \,  \lambda_i  \, 
[  \om  \, \widehat {\Gamma}_i \tau_2(t) - \tau_2(t) \widehat {\Gamma}_i  ]  \period  \ee
Here $ i = 1, \ldots , NL$. 

The characteristic polynomial of $H$ is
\bd | H- \lambda I | \; = \; s_0  \, \lambda^{NL} + s_1 \, \lambda^{N(L-1)}+ s_2 \,  
 \lambda^{N(L-2)} +   \, \cdots \, + s_{L-1} \,   \lambda^{N} + s_L  \comma \ed
  which is the RHS of (\ref{eqr}).
   The eigenvalues of $H$ are therefore  $\lambda_{p,k} 
 = \om^{p} \,  r_k $, where $p  = 0, 1, \ldots , N-1$ and 
 $ k=1, \ldots , L$ as in (\ref{res}).
 We can therefore naturally replace $i$ in (\ref{eqsh}), (\ref{eqst}) by 
  the pair of numbers $p, k$.

The matrices $\cal H $, $\tau_2(t)$ are of dimension $N^L$. They
 commute (for all $t$), so there is a similarity transformation, independent of $t$,\
 that diagonalizes both. Write their elements in this representation as 
  ${\cal H}_m  \, \delta_{mn}$,
${{\taub}_m} \, \delta_{mn}$.
 
  %%34567890123456789012345678901234567890123456789012345678901234567890

  Let us go this representation. The $ \widehat{\Gamma}_i$ are also of dimension $N^L$, 
  but do not commute  with either $\cal H$ or $\tau_2(t)$, so are not diagonal in this 
 representation. Write their elements as $(\widehat{\Gamma}_{p,k})_{mn}$. Then 
 (\ref{eqsh}),  (\ref{eqst})  become
 \be  \label{eigh}
[{\cal H}_n - {\cal H}_m -(1-\om^{-1}) \om^p  \, r_k]  \,  
 (\widehat{\Gamma}_{p,k})_{mn}  =  0  \comma  \ee
 
  \be \label{eigt}
[(1- \om^{p+1} \,  t \,  r_k) \taub_n   - (1- \om^{p} \,  t \, r_k) \taub_m   ]  \,  
 (\widehat{\Gamma}_{p,k})_{mn}  =  0  \period  \ee
 
 {From} (\ref{defHH}), $ \taub_m = 1 + \om t {\cal H}_m + {\rm O} (t^2)$. 
 Expanding (\ref{eigt}) to first order in $t$, we obtain  (\ref{eigh}).
 
Let us use  the known values (\ref{res}), (\ref{eigH}) of the eigenvalues of 
 $\tau_2(t)$ and $\cal H$, we see that  (\ref{eigh}), (\ref{eigt}). 
 Write 
 \be m = \{ m_1, \ldots , m_L \} \sep n = \{ n_1, \ldots,  n_L \} \comma \ee
 where the $m_i, n_i$ take the values $0,1, \ldots, N-1$, ordering the
 elements so that (\ref{defH}) is equivalent to
 \be \label{eigh2}
 {\cal H}_m \eq -\sum_{j=1}^L   \om^{m_j} \, r_j  \ee
 and (\ref{res}) to
 \be  \label{tau22}
 \taub_m \eq \prod_{k=1}^L (1- \om^{m_k+1} r_k t ) \period \ee
 Then (\ref{eigh}), (\ref{eigt}) are satisfied if,  when 
 $ (\widehat{\Gamma}_{p,k})_{mn} $ is non-zero,
 \be \label{restr}
 m_k  \; = \;  n_k+1 \eq p \; \; \;  ({\rm mod} \; N), \; \; \; 
\; \; \; m_j   \; =  \;   n_j \; \; \; {\rm  for } \; \; j \neq k \period \nonumber  \ee
  This means that $X_k^{-1}  \,  \widehat{\Gamma}_{p,k}$ is a diagonal matrix, 
  of   rank  $N^{L-1}$.
 
 We observe that  the elements of $(\widehat{\Gamma}_{p,k})_{mn} $ are indeed
 non-zero when  (\ref{restr}) is satisfied. If we assume that this is so, then (\ref{eigh})
 defines the eigenvalues of $\cal H$ to within the addition of a single constant to all.
 {From} (\ref{hamH}), ${\rm trace} \, {\cal H} = 0$, which fixes this constant. It follows that
 the eigenvalues of  $\cal H$ are indeed given by (\ref{eigh2}), so Fendley's 
 method is indeed a new way of  obtaining them.
 
 Similarly (\ref{eigt}) gives  the eigenvalues of $\tau_2(t)$ to within an
 overall multiplicative factor. One way of determining this would be to use
(\ref{tz}) and the fact that $\tau_2(t)$ is a polynomial in $t$ of degree $L$.
This would necessarily give (\ref{tau22}).

\section{Parafermionic properties of the  $\Gamma_j$ }
\setcounter{equation}{0}
Fendley noted various relations between the $\widehat{\Gamma}_i$. A significant 
one is that if $G$ is any linear combination of the 
$\widehat{\Gamma}_{p,k}$, i.e.
\be G \eq \sum_{p = 0}^{N-1} \sum_{k=1}^L   \rho_{p,k} \, 
\widehat{\Gamma}_{p,k}   \ee
where  the coefficients   $\rho_{p,k}$ are arbitrary, then
$G^N$ is proportional to the identity matrix $I$. More precisely,
\be \label{paraferm}
G ^N \eq  \left(  \sum_{k=1}^L  \beta_k \, 
 \rho_{0,k} \cdots \rho_{N-1,k} \right)  \, I \ee
where $\beta_1, \ldots, \beta_L$ are scalar factors, independent of the 
$\rho_{p,k}$. {From}  (\ref{lintr})  each $\widehat{\Gamma}_{p,k}$ is a
 linear combination of the $\Gamma_j$, so it is also true
  that the $N$th power of any  linear combination of the
$\Gamma_j$ matrices is proportional to the identity matrix. This is a 
natural extension of the anti-commutation property of the free-fermion or Clifford algebra 
used by Kaufman\cite{Kaufman49} for the Ising model.

%%34567890123456789012345678901234567890123456789012345678901234567890

There are also  quadratic relations between the $\widehat{\Gamma}_i$. As above, we 
write $\widehat{\Gamma}_i$ as 
 $\widehat{\Gamma}_{p,k}$.  Then
 \be  \label{rel1}
 \widehat{\Gamma}_{p,k} \, \widehat{\Gamma}_{p',k}  \eq 0 \; \; 
{\rm if } \; \;   p'  \neq  p-1,  \; \; {\rm mod}  \; N \, ,  \ee
and, for all $p,  k,  p',  k'$,
\be  \label{rel2}
  (r_k \,  \om^p - r_{k'} \, \om^{p'
  + 1} )   \,  \widehat{\Gamma}_{p,k}  \,   \widehat{\Gamma}_{p',k'}  +  ( r_{k'} 
  \, \om^{p' }  - r_k  \, \om^{p + 1} ))
\,   \widehat{\Gamma}_{p',k'}  \,   \widehat{\Gamma}_{p,k}  \eq 0 \period  \ee

The relation (\ref{rel1}) implies  that both sides of (\ref{rel2}) are zero if $k = k'$, for all
$p, p'$. For $N=2,\ldots, 6$, we have verified that (\ref{rel2}) implies 
(\ref{paraferm}).

\section{Group property of the $\tau_2(t)$ matrices }
\setcounter{equation}{0}

Going back to equation (\ref{treps}), consider the set of all $N^L$-dimensional 
invertible  matrices $V$ such that $V \Gamma_j V^{-1}$ is a linear combination 
of $\Gamma_0, \ldots, \Gamma_{q}$, i.e.  $\exists$ an $NL$ by $NL$ matrix  $v$ 
with elements  $v_{jk}$ such that
\be \label{groupV}
V \, \Gamma_j \, V^{-1}  \eq \sum_{k=0}^q v_{jk} \Gamma_k \period \ee
Such a set is necessarily a group $\cal G$. If $V$ can be arbitrarily close to the identity $I$, 
then
we can write
\be V = I   +   \epsilon {\widetilde{\cal H} } + O(\epsilon^2) \comma \ee
expanding (\ref{groupV}) to first order in $\epsilon$, we obtain a generalization of 
(\ref{Heqn}):
\be  \label{Heq2}
{\widetilde{\cal H}} \Gamma_j - \Gamma_j {\widetilde{\cal H}}  \eq 
\sum_{k=0}^q \tilde{h}_{jk}  \Gamma_k \period \ee
 
 Given the $\Gamma_j$, this is a linear equation for ${\widetilde{\cal H}}, \tilde{h}_{jk}$.
 For $N >2$, if we work
 in the representation where $\tau_2(t), { \cal H} $ are diagonal, then 
 we find numerically that the only 
 solutions for $\widetilde{\cal H}$, $\tilde{h}$   of  (\ref{Heq2}) are also diagonal. Thus in 
any representation the only solutions  commute with $\tau_2(t)$ and this in turn 
suggests that  the only solutions of  (\ref{groupV})  may be matrices $V$ that 
commute with $\tau_2(t)$. This is {\em not}
 true for $N=2$, where  $\log V, {\cal H}$  can be arbitrary  quadratic forms in the
 $2L$  $\Gamma$'s: such forms do not in general all commute. In this sense
 the group $\cal G$ is more restricted when $N>2$ than when $N=2$.

\section{Summary}
We have indicated how Fendley's method can be generalized from 
the simple hamiltonian (\ref{hamH2}) to the more general hamiltonian 
(\ref{hamH}), and to the corresponding $\tau_2$ model with open boundaries. We 
have defined a set of operators  $\Gamma_0, \ldots, \Gamma_{NL-1}$
that satisfy the commutation relations (\ref{Heqn}) with $\cal H$,
as well as the relations  (\ref{Heqn2}), (\ref{treps})  with $\tau_2(t)$.
They have the parafermionic properties (\ref{paraferm}) - (\ref{rel2}) observed by 
Fendley. 

We emphasize again that most of the equations in sections 4, 5 and 6 were
conjectured by the author,  based on the results of computer algebra calculations 
for small $N$ and $L$. Proofs of (\ref{relngammas}) and (\ref{numu}) 
are given  in the following paper.\cite{AuYangPerk}

 A significant difference from the usual Clifford algebra is that  the $\Gamma$
 matrices depend on the hamiltonian $\cal H$, or on the transfer matrix $\tau_2(t)$.
 Also, as discussed in section 6, for $N > 2$ the group $\cal G$ appears to be 
 restricted to  matrices that commute with $\tau_2(t)$.
 Even so, as Fendley says, this method opens up some very interesting possibilities.
 In particular, can it be extended to an hamiltonian that has some kind of cyclic 
 property?\footnote{If one simply extends the second 
 summation in (\ref{hamintro}) to include $j = L$, taking 
 $Z_{L+1}$ to be $Z_1$, one does not appear to obtain an eigenvalue spectrum 
 with the simple form (\ref{dirsum}), but perhaps there is another way of 
 making the extension that does preserve that simple ``direct sum''  form of 
 the eigenvalue spectrum.}
 
 The solvable chiral Potts model has a superintegrable case that is the $\tau_N$ model
 turned through $90^{\circ}$. Does this parafermion algebra play a useful role
 in that more general model? The corner transfer matrix  method of calculating order 
 parameters works well for solvable two-dimensional models with the 
 rapidity-difference property, but seems to  utterly fail for the chiral Potts 
 model.\cite{Bax06}
 Does this parafermion algebra provide a way of working with $NL$-dimensional 
 representatives instead of the full $N^L$-dimensional corner transfer matrices, as 
 can be done  for the Ising case $N=2$?\cite{Bax77}
 
 \section{Acknowledgement}
 
 The author is grateful to Paul Fendley for sending 
 him his paper,  and to Helen Au-Yang and Jacques Perk for valuable correspondence on this topic.

 %%34567890123456789012345678901234567890123456789012345678901234567890

\end{document}